\documentclass[pre,aps,showpacs,epsf,preprint]{revtex4}
\usepackage{amsmath,amssymb}
\usepackage{graphicx}
\usepackage{psfrag}
\usepackage{graphicx}

\begin{document}
\sloppy \draft
\title{Fluctuating hydrodynamics and turbulence in a rotating fluid: Universal
properties}

\author{Abhik Basu}
\email{abhik.basu@saha.ac.in} \affiliation{ Theoretical Condensed
Matter Physics Division, Saha Institute of Nuclear Physics, 1/AF
Bidhannagar, Kolkata (Calcutta) 700 064, India}
\author{Jayanta K Bhattacharjee}
\email{jkb@bose.res.in} \affiliation{ Satyendranath Bose Centre for
Basic Sciences, Sector III, Saltlake, Kolkata (Calcutta) 700 098,
India}

\begin{abstract}
We analyze the statistical properties of three-dimensional ($3d$)
turbulence in a rotating fluid. To this end we introduce a
generating functional to study the statistical properties of the
velocity field $\bf v$. We obtain the master equation from the
Navier-Stokes equation in a rotating frame and thence a set of exact
hierarchical equations for the velocity structure functions for
arbitrary angular velocity $\mathbf \Omega$. In particular we obtain
the {\em differential forms} for the analogs of the well-known von
Karman-Howarth relation for $3d$ fluid turbulence. We examine their
behavior in the limit of large rotation. Our results clearly suggest
dissimilar statistical behavior and scaling along directions
parallel and perpendicular to $\mathbf \Omega$. The hierarchical
relations yield strong evidence that the nature of the flows for
large rotation is not identical to pure two-dimensional flows. To
complement these results, by using an effective model in the
small-$\Omega$ limit, within a one-loop approximation, we show that
the equal-time correlation of the velocity components parallel to
$\mathbf \Omega$ displays Kolmogorov scaling $q^{-5/3}$, where as
for all other components, the equal-time correlators scale as
$q^{-3}$ in the inertial range where $\bf q$ is a wavevector in
$3d$. Our results are generally testable in experiments and/or
direct numerical simulations of the Navier-Stokes equation in a
rotating frame.
\end{abstract}

\maketitle

\section{Introduction}
Near the second order phase transition equilibrium systems exhibit
scaling behavior for thermodynamic functions and correlations. These
are characterized by certain scaling exponents which depend on the
spatial dimension $d$ and the symmetry of the order parameter
characterizing the phase transition. These, however, do not depend
on the parameters specifying the Hamiltonian \cite{fisherrev}. Time
dependent correlation functions, characterized by dynamic scaling
exponents also show similar universality \cite{hal}. These standard
universal properties of equilibrium critical dynamics are fairly
robust with respect to perturbations violating detailed
balance~\cite{tauber-etal:02}. In contrast, truly nonequilibrium
systems, like fluid and magnetohydrodynamic turbulence, surface
growth etc., are described by appropriate equations of motion and
exhibit much richer universal behavior. Non-equlibrium systems tend
to be more sensitive on the parameters that appear in the equations
of motion. For example, one finds that for the Kardar-Parisi-Zhang
equation, anisotropic perturbations are relevant in $d > 2$ spatial
dimensions, leading to rich phenomena that include novel
universality classes and the possibility of first-order phase
transitions and multicritical behavior~\cite{tauber_frey:02}.

Turbulence in fluid, described by the Navier-Stokes
equation~\cite{land,frish} for the evolution of the velocity field
$\bf v$, is a good candidate of systems out of equilibrium due to
the external drive acting on the system. Statistically steady fluid
turbulence in three- ($3d$) and two- ($2d$) dimensions show markedly
different behavior: In $3d$, homogeneous and isotropic turbulence is
characterized by a set of multiscaling exponents for the structure
functions (see below) for distance $r$ in the inertial range between
the forcing scale $L$ and the dissipation scale $\eta_d$ (i.e.,
$\eta_d\ll r\ll L$), and forward cascade (from small to large
wavenumbers) of the energy. Turbulence in $2d$ shows an inverse
cascade of kinetic energy from the energy-injection scale to larger
length scales and a direct cascade in which the enstrophy cascades
towards small length scales \cite{backward}; in many physical
realizations of $2d$ turbulence, there is an air-drag-induced
friction. In this direct-cascade regime, velocity structure
functions show simple scaling but their vorticity counterparts
exhibit multiscaling \cite{2d}, with exponents that depend on the
friction. In a rotating fluid isotropic symmetry is broken by the
global rotation. How the breakdown of rotational invariance affects
the universal properties remains a very important theoretical
question. These studies are also important for geophysical flows,
e.g., flows in ocean and atmosphere. Ref.~\cite{smith}, in numerical
simulations of forced rotating incompressible turbulence within a
periodic box of small aspect ratio, showed that above a critical
rotation, $3d$ forcing leads to a $2d$ inverse cascade. Further
Ref.~\cite{jfm}, in a spectral approach to rotating turbulence
applied to a specific eddy damped quasi-normal Markovian model,
showed a trend towards two dimensional behavior in presence of
rotation. Ref.~\cite{waleffe} in a helical decomposition of the
Navier-Stokes equation demonstrated similar trends of two
dimensionalization of energy transfer. In recent studies using a
shell model for rotating turbulent fluid, the authors showed how
quasi two-dimensional behavior emerge as rotation speed $\Omega$
increases \cite{sagar}. Recent Direct Numerical Study (DNS)
\cite{mueller} for rotating fluid turbulence suggest that the
 energy spectra for the velocity components perpendicular to $\bf
\Omega$ scales as $q_\perp^{-2}$ in the inertial range, with
wavevector $\bf q_\perp$ being perpendicular to $\mathbf\Omega$. It
is not fully established whether the statistical nature of the flow
for large-$\mathbf\Omega$ is truly two-dimensional or not. In this
context, Ref.~\cite{galtier} in a helical decomposition argued that
at the lowest order rotating turbulence is not the same as $2d$
turbulence.

Since a complete description of fully developed homogeneous and
isotropic $3d$ turbulence requires enumeration of all the
multiscaling exponents, it is important to find out the nature of
multiscaling in the presence of rotation. Until now, there is no
theoretical perturbative calculational framework to obtain these
multiscaling exponents within controlled perturbative
approximations. However, the multiscaling exponent for the
third-order structure function for the longitudinal component of the
velocity field is well-known: Defining $\Delta {\bf v} (x)= {\bf
v}({\bf x_1 +x}) - {\bf v}({\bf x_1})$, one has in the inertial
range
\begin{equation}
\langle [\Delta v(|{\bf r}|)\cdot {\bf \hat r}]^3\rangle =
-\frac{4}{5}\epsilon r,\label{vonkar}
\end{equation}
where $\epsilon$ is the total energy dissipation (see below) and
${\bf \hat r}={\bf r}/ r$. This is the well-known von Karman-Howarth
4/5-law of $3d$ fluid turbulence \cite{kolmo}. The corresponding
differential form is given by
\begin{equation}
\nabla_j \langle \Delta v^2 \Delta
v_j\rangle=4\epsilon.\label{diffkar}
\end{equation}
Subsequently, an analog of Eq.~(\ref{diffkar}), involving mixed
correlation tensor of the velocity and vorticity, has been obtained
for helical turbulence \cite{heli}. In particular they find
correlation
\begin{equation}
\langle v_i({\bf x})v_j({\bf x})w_j({\bf x} + {\bf
r})\rangle=-\frac{\overline\eta}{10}r_i,\label{helkar}
\end{equation}
It remains to find out analogs of these relations as above  in the
presence of rotation, which introduces helicity in the system.

Our main results in this article are (obtained by using
two different strategies):
\begin{itemize}
\item By using the Navier-Stokes equation in a rotating frame,
we set up the master equation for the generating functional of the
probability distribution of the velocity field differences. We use
this to obtain the exact hierarchical relations between different
order structure functions of velocity components. In particular we
obtain a set of {\em differential forms} for the analog of the
well-known von Karman-Howarth relation for non-rotating $3d$ fluid
turbulence. These relations are, however, non-integrable and involve
contributions from pressure, unlike non-rotating $3d$ turbulence.
Thus we find that, unlike the case of fluid turbulence in an
inertial frame, there are no closed equations for the third order
structure functions and hence no simple analog of the well-known von
Karman-Howarth relation for inertial frame fluid turbulence in
rotating turbulence. A combination of these relations yield a
compact differential form of von Karman-Howarth like relation, which
although identical to its isotropic analog cannot be integrated
owing to the underlying anisotropy. We examine the relations in the
limit of large-$\Omega$. In this limit, we obtain simple relations
connecting velocity structure functions with mixed structure
functions, constructed out of gradients of pressure differences and
components of velocity differences. Finally, our results here are
illustrative of the differences between the statistical properties
of $3d$ turbulent flows in the presence of large rotation and those
of pure $2d$ turbulence.
\item From the Navier-Stokes equation in presence of rotation we
define an {\em effective model} which is expected to be valid in the
small-$\Omega$ limit. We apply mode-coupling methods on this
effective model to calculate the equal-time two-point correlation
functions of the velocity field components $v_i,\,i=x,y,z$. We find
that for $i=z$, $\langle|v_z({\bf q},t)|^2|\rangle\sim q^{-11/3}$,
where as $\langle v_i ({\bf q},t)v_j ({\bf -q},t)\rangle\sim
q^{-3},\,i,j\neq z$, where $\mathbf\Omega$ is along the
$z$-direction and $\bf q$ is a $3d$ wavevector. Thus, correlations
involving $v_x$ or $v_y$ exhibit scaling $\sim q^{-3}$ demonstrating
the anisotropic nature of rotating turbulence.
\end{itemize}
The remaining part of the paper is organized as follows: In
Sec.~\ref{eom} we write down the Navier-Stokes equation in the
presence of a global rotation. In Sec.~\ref{hierar} we set up the
hierarchical relatione between equal-time structure functions of
different order and the effects of rotation on them. We then define
our {\em effective model} to study rotating turbulence in
Sec.~\ref{perturb}. We use it to perform a one-loop self-consistent
(OLSC) calculation on the effective model to calculate various
equal-time velocity two-point correlation functions and enumerate
their scaling in the inertial range. We examine the limit of
large-$\mathbf\Omega$. Finally, we summarize and discuss our results
in Sec.~\ref{conclu}.

\section{Equation of motion}
\label{eom}

The Navier-Stokes equation in presence of a global rotation
${\mathbf \Omega}=\Omega \hat z$ is given, in the rotating frame, by
\begin{equation}
\frac{\partial {\bf v}} {\partial t}+ 2({\mathbf \Omega \times \bf
v}) +{\bf v}.\nabla {\bf v}=-{\nabla p^*\over \rho} +\nu_0 \nabla^2
{\bf v} +{\bf f} \label{navier}
\end{equation}
with $\nabla {\bf .v}=0$ (conclusion of incompressibility). In
Eq.(\ref{navier}), $p^*$ is the effective pressure which includes
the centrifugal force (through an effective potential
$\frac{1}{2}|\mathbf\Omega\times\bf r|^2$), $\nu_0$ is the kinematic
viscosity and $\bf f$ is a large-scale external forcing function,
required to maintain a statistical steady-state. The term
$2({\mathbf \Omega} \times\bf v)$ is the Coriolis force. Note that
Eq.~(\ref{navier}) is invariant under
\begin{equation}
z\rightarrow -z, \, v_z \rightarrow -v_z.\label{symm1}
\end{equation}
However, the symmetry under ${\bf r}_\perp \rightarrow -{\bf
r}_\perp,\, {\bf v}_\perp \rightarrow - {\bf v}_\perp$, which is
present in the Navier-Stokes equation without rotation, is broken in
Eq.~(\ref{navier}) due to rotation, where ${\bf r}_\perp = (x,y),\,
{\bf v}_\perp = (v_x, v_y)$. Thus rotation breaks the invariance
under in-plane parity inversion.  In a nonequlibrium system there is
no particular relation between the noise variance and the
dissipation coefficient, unlike in equilibrium systems where such a
relation exists due to the Fluctuation Dissipation Theorem
\cite{chai}. A promising starting point for theoretical/analytical
studies on homogeneous and isotropic fluid turbulence is the
randomly forced Navier-Stokes model, where the forcing function $\bf
f$ is a Gaussian random force whose
spatial Fourier transform ${\bf f} ({\bf q},t)$ has zero mean and %a variance
%We take $\bf f$ to be a zero-mean, Gaussian noise with
covariance \cite{yakhot}
\begin{equation}
\langle f_i({\bf q},\omega)f_j({\bf -q}, -\omega)\rangle=P_{ij}({\bf
q}){2D_0\over q^{d-4+y}} \label{yakhotnoise}
\end{equation}
 in $d$-dimensions where
$P_{ij}=\delta_{ij}-\frac{q_iq_j}{q^2}$ is the transverse projection
operator in the Fourier space, and $\bf q$ and $\omega$ are wave
vector and frequency respectively. This was used in
Ref.~\cite{yakhot} to calculate various universal quantities
associated with $3d$ turbulence. The Model A (with $y=2-d$) and
Model B (with $y=4-d$) of Ref.~\cite{fns} may be considered as
special cases of (\ref{yakhotnoise}).  In absence of any global
rotation ($\bf \Omega =0$) in $3d$, for $y=4$ one obtains
 the famous Kolmogorov (K41) result for the energy spectrum:
$E(q)=K_o\epsilon ^{2/3}q^{-5/3}$ for wavenumber $k$ in the inertial
range. Here $K_o$ is the Kolmogorov's constant (a dimensionless
universal number) and $\epsilon$ is energy dissipation rate per unit
mass which is the flux of the energy in the steady state. In a
rotating fluid system there are two important dimensionless numbers,
namely, the Rossby number $Ro=V/(2\Omega L)$ and the Ekman number
$M_0=\nu_0/(2\Omega L^2)$ where $V$ and $L$ are the characteristic
velocity and length scales, respectively. From Eq.(\ref{navier}) it
is clear that the possible modifications of the statistical
properties due to the rotation is essentially a non-linear effect,
because without the nonlinearity, the rotation only causes the $x-$
and the $y-$components of $\bf v$ to rotate. From a statistical
mechanics point of view, quantities of interests are the
correlations $C_{ij}(q,\omega)=\langle v_i({\bf q},\omega) v_j({-\bf
q},-\omega)\rangle$. For homogeneous and isotropic turbulence (i.e.,
for ${\mathbf \Omega}=0$), $C_{ij}(q,\omega) \sim P_{ij}({\bf q})
q^{-2\chi-z-3}g(\omega/q^z)$
 for wavevector $\bf q$ in the inertial range, $\chi$ and $z$ are the roughness
and the dynamic exponents respectively, $g$ is a scaling function.
Due to the Galilean invariance of Eq.(\ref{navier}) $\chi+z=1$
exactly \cite{yakhot}. It remains to be seen how a finite rotation
affects the scaling obtained in the isotropic situation.

\section{Hierarchical relations between structure functions}
\label{hierar}

In this Section we set up the exact hierarchical equations between
the two-point structure functions of various orders by generalizing
the generating-functional method of Ref.~\cite{polyakov}. These
equations are however not closed. These exact relations,
nevertheless, reveal explicitly the differences between the rotating
and the non-rotating cases.

In order to find the hierarchy of equations for the equal-time
structure functions in the statistical stationary state of the
stochastically forced Navier-Stokes equation in  a rotating frame
for the velocity field $v_i$:
\begin{eqnarray}
{\mathcal Z}({\mathbf \lambda}_1, {\mathbf \lambda}_2, {\bf r}_1,
{\bf r}_2, t)&\equiv& \langle \exp ({\mathbf \lambda}_1\cdot {\bf
v}({\bf x}_1) + {\mathbf \lambda}_2\cdot {\bf v}({\bf x}_2)\rangle=
\langle {\mathcal Z}_0\rangle \nonumber \\
&=&\int {\mathcal D}v_1 {\mathcal D}v_2{\mathcal P} ({\bf v}_1, {\bf
v}_2, {\bf x}_1, {\bf x}_2,t) \exp [{\mathbf \lambda}_1\cdot {\bf
v}({\bf x}_1) + {\mathbf \lambda}_2\cdot {\bf v}({\bf x}_2)],
\end{eqnarray}
where $\mathcal P$ is the joint probability distribution of
velocities ${\bf v}_1 ({\bf x}_1, t), {\bf v}_2 ({\bf x}_2,t)$. The
Eq. of motion for $\mathcal Z$ may be derived in a straightforward
way:
\begin{eqnarray}
\frac{\partial {\mathcal Z}}{\partial t} &=& -\frac{\partial^2
{\mathcal Z}}{\partial \lambda_j^\mu\partial x_j^\mu}-2\lambda_j^\mu
\epsilon_{jps}\frac{\partial {\mathcal Z}}{\partial
\lambda_s^\mu}\Omega_p + I_p + I_f +D,\label{eqzbasic}
\end{eqnarray}
where $D=\langle \nu_0 [{\mathbf \lambda}_1\cdot \nabla^2 {\bf
v}({\bf x}_1) + {\mathbf \lambda}_2 \cdot \nabla^2 {\bf v}({\bf
x}_2)]{\mathcal Z}_0\rangle$ are the anomaly terms, $I_p= \langle
[{\mathbf \lambda}_1 \cdot \nabla p^* ({\bf x}_1) + {\mathbf
\lambda}_2\cdot \nabla p^*({\bf x}_2)]{\mathcal Z}_0\rangle$ are the
pressure contributions, and $I_f=\langle [{\mathbf \lambda}_1\cdot
{\bf f}({\bf x}_1) +{\mathbf \lambda}_2\cdot {\bf f}({\bf
x}_2)]{\mathcal Z}_0\rangle$. Note that the master equation
(\ref{eqzbasic}) is not closed because of the anomaly terms $D$.
This nontrivial point has been treated by a variety of approaches
ranging from approximate techniques to rigorous studies
\cite{polyakov,gurarie-migdal}. The problem arises when we look at
the master equation in the limit of vanishing viscosity
$\nu_0\rightarrow 0$; here the anomaly terms produce a finite
effect. For instance, the finiteness of the dissipation in the limit
of vanishing viscosity (discussed above) is just produced by the
anomaly terms $D$. In what follows we shall specify the effects of
the anomaly terms in a more detailed way. Further, we can apply the
Furutsu-Novikov-Donsker \cite{Woy,McComb} formalism to calculate the
random forcing terms. For Gaussian distributed, white-in-time random
forces ${\bf f}$, which are additive terms in Eq.~(\ref{navier}), we
obtain the the final form of {\em the master equation} for
equal-time, two-point generating functions ${\mathcal Z}$:
\begin{eqnarray}
\frac{\partial {\mathcal Z}}{\partial t} &=& -\frac{\partial^2
{\mathcal Z}}{\partial \lambda_j^\mu\partial x_j^\mu}-2\lambda_j^\mu
\epsilon_{jps}\frac{\partial {\mathcal Z}}{\partial
\lambda_s^\mu}\Omega_p + I_p + D+ [K(0) - K(r)]{\mathcal Z},
\label{eqzbasic1}
\end{eqnarray}
where $K(r)$ is given by $\langle {\bf v}(r,t)\cdot{\bf
f}(0,t)\rangle$ (to be calculated by using the
Furutsu-Novikov-Donsker formalism), which is related to the variance
of ${\bf f}$.

In order to proceed further it is useful to apply the basic
symmetries of the dynamical equations to simplify the structure of
the master equation.  We assume that statistically stationary
turbulence has been produced under the dynamics of the
stochastically forced Navier-Stokes Eq.~(\ref{navier}). Stationarity
implies
\begin{eqnarray}
\partial_t{\mathcal Z}=0.
\end{eqnarray}
The Navier-Stokes equation without rotation (i.e., Eq.(\ref{navier})
with $\mathbf \Omega=0$) is invariant under uniform translation. For
$\mathbf\Omega\neq 0$, the presence of the centrifugal force
$\mathbf \Omega \times (\mathbf\Omega\times \bf r)$, which is
contained inside effective pressure $p^*$, breaks the translational
invariance. Since we are considering incompressible flows, pressure
$p^*$ may be eliminated by using the condition $\nabla\cdot{\bf
v}=0$. We obtain
\begin{equation}
-\nabla^2 p^*= \nabla\cdot ({\mathbf \Omega\times \bf v}) + \nabla_j
({\bf v}\cdot \nabla) v_j.\label{press}
\end{equation}
The resulting equation after eliminating $p^*$ then becomes
\begin{equation}
\frac{\partial v_i}{\partial t} + 2P_{ij}({\mathbf \Omega\times\bf
v})_j + P_{ij} {\bf v}\cdot \nabla v_j = \nu_0 \nabla^2 v_i +
f_i.\label{incomnavier}
\end{equation}
Evidently, Eq.~(\ref{incomnavier}) is invariant under ${\bf
r\rightarrow r} + {\bf r}_0$ where ${\bf r}_0$ is a constant vector.
Thus for the hierarchical relations between structure functions for
an incompressible fluid, we impose homogeneity on $\mathcal Z$:
\begin{eqnarray} {\mathcal Z}={\mathcal
Z}({\mathbf\lambda}_1, {\mathbf\lambda}_2, {\bf x}_1-{\bf x}_2).
\end{eqnarray}
Equivalently, if we define
\begin{eqnarray}
{\bf x}^+\equiv {\bf x}_1+ {\bf x}_2,\;\;{\bf r}\equiv {\bf x}_1 -
{\bf x}_2,
\end{eqnarray}
we can write
\begin{eqnarray}
\frac{\partial}{\partial x_{1i}}{\mathcal Z}=
\frac{\partial}{\partial x_{+i}}{\mathcal Z}+
\frac{\partial}{\partial r_i}{\mathcal Z}, \label{eq:eq2.27}
\\
\nonumber\\
\frac{\partial}{\partial x_{2i}}{\mathcal Z}=
\frac{\partial}{\partial x_{i+}}{\mathcal Z}-
\frac{\partial}{\partial x_{ri}}{\mathcal Z}. \label{eq:eq2.28}
\end{eqnarray}
Homogeneity now implies $({\partial}/{\partial x_{i+}}){\mathcal
Z}=0$.

The $3d$ Navier Stokes Equation (\ref{navier}) is invariant under
the Galilean transformation
\begin{eqnarray}
{\bf x}={\bf x'}+{\bf u}_0t',\,\,\,t=t',\,\,\,{\bf v}({\bf x},
t)={\bf v'}({\bf x'}, t')+{\bf u}_0.
\end{eqnarray}
Since ${\mathcal Z}=\langle \exp ({\mathbf \lambda}_1\cdot {\bf
v}({\bf x}_1) + {\mathbf \lambda}_2\cdot {\bf v}({\bf x}_2)\rangle$,
this Galilean invariance implies
\begin{eqnarray}
{\mathcal Z}={\mathcal Z}({\mathbf \lambda}_1-{\mathbf \lambda}_2,
{\bf x}_1 - {\bf x}_2).
\end{eqnarray}
If we now introduce variables
\begin{eqnarray}
{\mathbf \lambda}_0&=&{\mathbf \lambda}_1 +{\mathbf
\lambda}_2,\nonumber \\
{\mathbf \lambda}&=&{\mathbf \lambda}_1 - {\mathbf \lambda}_2,
\end{eqnarray}
and use the same considerations as in Eqs. (\ref{eq:eq2.27}) and
(\ref{eq:eq2.28}) for ${\bf x}_1,\,{\bf x}_2$, we see that the
Galilean invariance is equivalent to demanding
$\frac{\partial}{\partial \lambda_{1i}}{\mathcal Z}=0=
\frac{\partial}{\partial \lambda_{2i}}{\mathcal Z}$. Therefore the
master equation can be conveniently rewritten in terms of the
variables ${\mathbf \lambda}$ and $\bf r$ to get %the master equation
%for
the generating function
\begin{eqnarray}
{\mathcal Z}({\bf r}, \,{\mathbf \lambda}) &=&\langle \exp[{\mathbf
\lambda}\cdot {\bf \Delta v}]\rangle\equiv \langle {\mathcal
Z}_0\rangle,\;\;{\textrm ({\rm with}}\;\,{\bf \Delta v}={\bf v}({\bf
x}_1) - {\bf v}({\bf x}_2)\,{\textrm )},\label{zdef}
\end{eqnarray}
which obeys the following equation in steady-state:
\begin{eqnarray}
\frac{\partial^2 {\mathcal Z}}{\partial \lambda_j\partial r_j} +
4\lambda_j \epsilon_{jzs}\frac{\partial {\mathcal Z}}{\partial
\lambda_s}\Omega = I_p + I_f +D,\label{zfinaleq}
\end{eqnarray}
where we have used ${\mathbf \Omega}=\Omega \hat z$. Due to the
rotation (about the $z$-axis), the system is anisotropic and $3d$
rotational invariance is broken. However the system admits {\em
restricted two-dimensional rotational invariance} about the
$z$-axis. This dictates that $\mathcal Z$ depends separately on $z$
and ${\bf r_\perp}$ ($\bf r_\perp$ lies in the $xy$-plane). We
define ${\mathbf \lambda}=(\lambda_z,\,\eta_1,\,\eta_2)$, i.e.,
$\lambda_z,\,\eta_1,\,\eta_2$ are the $z,\,r$ and
$\theta$-components of $\mathbf \lambda$, respectively. In terms of
these variables $\mathcal Z$ may be expressed as
\begin{equation}
{\mathcal Z}=\langle \exp [\lambda_z \Delta v_z (z,{\bf r_\perp})+
\eta_1\Delta v_r (z,{\bf r_\perp}) + \eta_2\Delta v_\theta (z,{\bf
r_\perp})]\rangle, \label{zdefi}
\end{equation}
where $\Delta v_z,\Delta v_r$ and $\Delta v_\theta$ are $z,r$ and
$\theta$ components of the velocity difference vector $\bf \Delta
v$. The anomaly terms $D$ may be written in terms of the energy
dissipation rate per unit mass $\epsilon$ as given below (see
Appendix for details)
\begin{eqnarray}
D&=&-2\nu_0\langle[\lambda_z^2 (\nabla_i v_z)^2 + \eta_1^2(\nabla_i
v_r)^2 +\eta_2^2 (\nabla_iv_\theta)^2 + 2\lambda_z\eta_1 (\nabla_i
v_z)(\nabla_i v_r) \nonumber \\ &+& 2\eta_1\eta_2(\nabla_i
v_r)(\nabla_i v_\theta) + 2\eta_2\lambda_z (\nabla_i
v_\theta)(\nabla_i v_z)]{\mathcal Z}_0\rangle. \label{anomaly}
\end{eqnarray}
In a statistical steady state, these terms are nonzero constants and
are related to the energy input that is determined by the variances
of the external stochastic forces. We are interested in the scaling
properties of various structure functions in the inertial range
($r\ll L=1$, where $L$ is the system size), where the force
correlator $K(r)$ may be expanded in powers of $r$. To the leading
order we assume the following Taylor expansion: $K(r)=K(0)-Ar^2$,
where $A$ is a numerical constant. We then neglect the $r^2$ term
relative to contributions from $D$. We now proceed to calculate
hierarchical relations between structure functions of different
orders. It is seen easily that differentiation with respect to
$\lambda_z,\,\eta_1,\,\eta_2 $ leads to various structure functions
of different order: We write $S_{m,n,p}=\langle (\Delta
v_z)^m(\Delta v_r)^n (\Delta
v_\theta)^p\rangle=\frac{\partial^{m+n+p}}
{\partial\lambda_z^m\partial\eta_1^n\partial\eta_2^p}{\mathcal
Z}|_{\lambda_z=0,\eta_1=0,\eta_2=0}$. Now $\mathcal Z$ as a function
of $\lambda_z,\,\eta_1,\,\eta_2$ as defined in (\ref{zdefi})
satisfies the steady-state equation
\begin{eqnarray}
&&\frac{\partial^2{\mathcal Z}}{\partial z\partial \lambda_z} +
\frac{1}{r_\perp}\frac{\partial}{\partial
r_\perp}\left(r_\perp\frac{\partial {\mathcal Z}}{\partial
\eta_1}\right)+\frac{1}{r_\perp}\left(-\eta_2
\frac{\partial}{\partial\eta_1} + \eta_1\frac{\partial}
{\partial\eta_2}\right)\frac{\partial \mathcal Z}{\partial \eta_2}
+4\Omega \left(-\eta_1 \frac{\partial}{\partial\eta_2} +
\eta_2\frac{\partial}{\partial \eta_1}\right){\mathcal Z}\nonumber
\\
&=& I_p +I_f +D. \label{zfinaleqI}
\end{eqnarray}
It will now be useful to consider the pressure contribution: The
Navier-Stokes Equation in an inertial frame ($\Omega=0$) is
invariant under the parity transformation: $\bf r\rightarrow -r,\,
v\rightarrow -v$, where $\bf r$ is the $3d$ radius vector. %This yields for functions
%\begin{eqnarray}
%\langle [\nabla ({\bf x}_1) p^* -\nabla ({\bf x}_2)p^*](\Delta
%a)^s\rangle &=&0\;\;\;{\rm if} s=2n,\nonumber \\
%&=&\neq 0 \;\;\;{\rm if} s=2n+1.
%\end{eqnarray}
This  yields that there are no pressure contributions to the
well-known von Karman-Howarth relation for the third order
longitudinal velocity structure function in non-rotating turbulence
\cite{frish,yakhot1,sreeni}. When there is a finite rotation
($\Omega\neq 0$), there is no invariance under the parity
transformation above, and as a result, certain pressure
contributions are finite (see below) which are zero in the
non-rotating case. In addition, the effective pressure $p^*$ now
explicitly depends on $\Omega$ as given by Eq.~(\ref{press}), and
hence will be large in the large-$\Omega$ limit. With this
discussion in mind, let us now write down the hierarchical relations
between structure functions of various orders: Separately apply
$\frac{\partial}{\partial\lambda_z},\,\frac{\partial}{\partial\eta_1},\,
\frac{\partial}{\partial\eta_2}$ and setting
$\lambda_z,\,\eta_1,\,\eta_2$ to zero yield
\begin{eqnarray}
\frac{\partial}{\partial z}\langle (\Delta v_z)^2\rangle +
\frac{\partial}{\partial r_\perp} \langle \Delta v_r \Delta
v_z\rangle +
\frac{1}{r_\perp}\langle \Delta v_r\Delta v_z\rangle &=&0, \nonumber \\
\frac{\partial}{\partial z} \langle \Delta v_z \Delta v_r\rangle +
\frac{1}{r_\perp}\frac{\partial}{\partial r_\perp}\langle r_\perp (\Delta v_r)^2\rangle + \frac{1}{r_\perp}\langle (\Delta v_\theta)^2\rangle&=&0,\nonumber \\
\frac{\partial}{\partial z} \langle \Delta v_z\Delta v_\theta\rangle
+ \frac{\partial}{\partial r_\perp} \langle  \Delta v_r\Delta
v_\theta\rangle &=&0. \label{2ndorder}
\end{eqnarray}
Note the above relations between the second order structure
functions are independent of $\mathbf\Omega$. In order to calculate
similar relations between different third order structure functions,
we separately apply $\frac{\partial}{\partial \lambda_z^2},\,
\frac{\partial}{\partial\eta_1^2},$ and
$\frac{\partial}{\partial\eta_2^2}$ and set $\lambda_z,\,
\eta_1,\,\eta_2$ to zero. We obtain
\begin{eqnarray}
&&\frac{\partial}{\partial z} \langle (\Delta v_z)^3\rangle +
\frac{\partial}{\partial r_\perp}\langle (\Delta v_z)^2\Delta
v_r\rangle+ \frac{1}{r_\perp}\langle (\Delta v_z)^2 \Delta
v_r\rangle
= -4\langle\epsilon_z\rangle,\nonumber \\
&&\frac{\partial}{\partial z}\langle \Delta v_z (\Delta
v_r)^2\rangle +\frac{\partial}{\partial r_\perp}\langle (\Delta
v_r)^3\rangle +\frac{1}{ r_\perp}\langle (\Delta v_\theta)^2\Delta
v_r\rangle + \frac{1}{r_\perp}\langle (\Delta v_r)^3\rangle -4\Omega
\langle \Delta v_\theta \Delta v_r\rangle \nonumber \\&=& -2\langle[
\frac{\partial \Delta p^*}{\partial r_\perp} \Delta
v_r]\rangle - 4\langle\epsilon_r\rangle,\nonumber \\
&&\frac{\partial}{\partial z}\langle \Delta v_z (\Delta
v_\theta)^2\rangle +\frac{\partial}{\partial r_\perp}\langle \Delta
v_r (\Delta v_\theta)^2\rangle + 4\Omega \langle \Delta v_\theta
\Delta v_r\rangle \nonumber \\&=&
-4\langle\epsilon_\theta\rangle.\label{3rdorderI}
\end{eqnarray}
Here $\epsilon_z = \nu_0(\nabla_i v_z)^2,\,\epsilon_r=\nu_0
(\nabla_i v_r)^2,\,\epsilon_\theta=\nu_0(\nabla_i v_\theta)^2$,
pressure difference $\Delta p^*= p^*({\bf x}_1) - p^*({\bf x}_2)$.
Adding the three relations (\ref{3rdorderI}) we obtain
\begin{eqnarray}
&&\frac{\partial}{\partial z} \langle (\Delta v_z)^3\rangle +
\frac{\partial}{\partial r_\perp}\langle (\Delta v_z)^2\Delta
v_r\rangle+ \frac{1}{r_\perp}\langle (\Delta v_z)^2 \Delta
v_r\rangle \nonumber \\ &+& \frac{\partial}{\partial z}\langle
\Delta v_z (\Delta v_r)^2\rangle +\frac{\partial}{\partial r}\langle
(\Delta v_r)^3\rangle + \frac{1}{r_\perp}\langle (\Delta
v_r)^3\rangle \nonumber
\\ &+& \frac{\partial}{\partial z}\langle \Delta v_z (\Delta
v_\theta)^2\rangle +\frac{\partial}{\partial r_\perp}\langle \Delta
v_r
(\Delta v_\theta)^2\rangle +\frac{1}{r_\perp} \langle (\Delta v_\theta)^2\Delta v_r\rangle \nonumber \\
&=& -2\langle[\frac{\partial \Delta p^*}{\partial r_\perp} \Delta
v_r]\rangle  - 4\langle \epsilon\rangle. \label{von1}
\end{eqnarray}
 Here
$\langle\epsilon\rangle$ is the total energy dissipation rate per
unit mass: $\langle \epsilon\rangle=\langle\epsilon_z\rangle
+\langle\epsilon_r\rangle +\langle\epsilon_\theta\rangle$. Relation
(\ref{von1}) may be written in a closed form, by using
incompressibility of the fluid, as
\begin{equation}
 \nabla_j \left[\langle |{\Delta \bf v }|^2 \Delta v_j\rangle +
\langle \Delta p^* \Delta v_j\rangle\right] = -4
\langle\epsilon\rangle, \label{voncompact}
\end{equation}
where we have used that $\langle \frac{\partial \Delta p^*}{\partial
z} \Delta v_z\rangle=0$ due to the invariance of the Navier-Stokes
equation (\ref{navier}) under $z\rightarrow -z,\, v_z\rightarrow
-v_z$. Further, $|\Delta {\bf v}|^2 = (\Delta v_z)^2 + (\Delta
v_r)^2 + (\Delta v_\theta)^2$. Equation (\ref{voncompact}) is the
{\em differential} analog of the well-known von Karman-Howarth
relation of fluid turbulence without rotation \cite{frish}. This may
be simplified further by using incompressibility: One obtains
$\nabla_j \langle |{\Delta \bf v }|^2 \Delta v_j\rangle=0$,
which is same as in Ref.~\cite{galkar}.
Although it is identical to the von Karman-Howarth relation
(\ref{diffkar}), it cannot be integrated due to the anisotropy.
Thus, in that sense, a simple (integral) analog of the von
Karman-Howarth relation for rotating fluid turbulence does not
exist. Equation (\ref{voncompact}) is one of the main results of
this article.  When $\mathbf \Omega=0$, the pressure contributions
in Eq.~(\ref{3rdorderI}) vanish and consequently one obtains the
result for the nonrotating case. Three more relations, analogous to
(\ref{3rdorderI}) may be obtained by separately applying
$\frac{\partial}{\partial\lambda_z}
\frac{\partial}{\partial\eta_1},\,
\frac{\partial^2}{\partial\lambda_z
\partial\eta_2},\,\frac{\partial^2} {\partial\eta_1\partial\eta_2}$
on Eq.~(\ref{zfinaleqI}) and setting $\lambda_z,\,\eta_1,\,\eta_2$
to zero. We find
\begin{eqnarray}
&&\frac{\partial}{\partial z}\langle (\Delta v_z)^2\Delta v_r\rangle +
 \frac{\partial}{\partial r_\perp}\langle \Delta v_z (\Delta v_r)^2\rangle
 + \frac{1}{r_\perp}\langle \Delta v_z (\Delta v_r)^2\rangle +\frac{1}{r_\perp} \langle \Delta v_z (\Delta v_\theta)^2\rangle
 - 4\Omega \langle \Delta v_z\Delta v_\theta\rangle
 \nonumber \\ &=&
 -\langle[\frac{\partial \Delta p^*}{\partial z}\Delta v_r]\rangle -
 \langle[ \frac{\partial \Delta p^*}{\partial r_\perp}\Delta
 v_z]\rangle, \nonumber\\
&& \frac{\partial}{\partial z}\langle (\Delta v_z)^2\Delta
 v_\theta\rangle + \frac{\partial}{\partial r_\perp}\langle \Delta
 v_z\Delta v_r \Delta v_\theta\rangle + 4\Omega \langle \Delta
 v_z\Delta v_r\rangle\nonumber \\&=&-\langle[\frac{\partial \Delta p^*}{\partial z}\Delta
 v_\theta]\rangle,
 \nonumber\\
&& \frac{\partial}{\partial z}\langle \Delta v_z \Delta v_r\Delta
 v_\theta\rangle +\frac{\partial}{\partial r_\perp}\langle (\Delta
 v_r)^2 \Delta v_\theta\rangle + \frac{1}{r_\perp}\langle (\Delta v_\theta)^2\Delta
 v_r\rangle + 4\Omega [-\langle(\Delta v_\theta)^2 \rangle+ \langle
 (\Delta v_r)^2\rangle]\nonumber \\&=&-4\langle\epsilon_{r\theta}\rangle-\langle[\frac{\partial \Delta p^*}{\partial r}\Delta v_\theta]\rangle,\label{3rdorderII}
 \end{eqnarray}
 where $\epsilon_{r\theta}=\nu_0 (\nabla_j v_r)(\nabla_j v_\theta)$
 is non-zero in the presence of rotation, but zero when ${\mathbf
 \Omega}=0$ (due to the invariance under the in-plane parity
 inversion).
 Similar to Eqs.~(\ref{3rdorderI}), Eqs.~(\ref{3rdorderII}) do not
 decouple, have second order structure functions and pressure
 contributions. In particular, with the exception of the
two-point structure function (in our notation defined above)
$S_{2,0,0}\equiv\langle (\Delta v_z)^2\rangle$, all other two-point
velocity structure functions appear in the relations involving
third-order structure functions (\ref{von1}) and (\ref{3rdorderII}).
Despite the limitations of these relations, useful
 information may be obtained in the limit $\Omega\rightarrow \infty$ from relations
 (\ref{3rdorderI}) and (\ref{3rdorderII}).
 Note that, from Eq.~(\ref{press}), in the limit $p^*/\Omega$ does not vanish, and hence we obtain
 \begin{eqnarray}
 \langle \Delta v_\theta\Delta  v_r\rangle&=& \frac{1}{2\Omega}
 \langle (\frac{\partial}{\partial r} \Delta p^*)\Delta v_r\rangle=0,\nonumber \\
  \langle \Delta v_z\Delta v_\theta\rangle &=&\frac{1}{4\Omega}[\langle
 \frac{\partial \Delta p^*}{\partial z}\Delta v_r\rangle +\langle
 \frac{\partial \Delta p^*}{\partial r}\Delta v_z\rangle],\nonumber \\
 \langle \Delta v_z\Delta v_r\rangle &=& -\frac{1}{4\Omega}\langle
  \frac{\partial \Delta p^*}{\partial z}\Delta v_\theta\rangle ,\nonumber \\
 -\langle (\Delta v_\theta)^2\rangle + \langle (\Delta v_r)^2\rangle &=&
 -\frac{1}{4\Omega}\langle \frac{\partial \Delta p^*}{\partial r_\perp}\Delta
 v_\theta\rangle.
  \label{largeomega}
 \end{eqnarray}
 These
are {\em exact} relations between certain two-point velocity
structure functions and two-point mixed structure function of
pressure gradients and velocity differences in the limit of large
rotation. Thus we find that in this limit (i) $\langle \Delta
v_\theta \Delta v_r\rangle=0$, (ii) second order structure functions
$\langle\Delta v_z \Delta v_\theta\rangle$, $\langle\Delta v_z
\Delta v_r\rangle$ and the combination $\langle (\Delta
v_r)^2\rangle - \langle (\Delta v_\theta)^2\rangle$ are simply
related to certain mixed structure functions involving pressure
difference gradient and velocity difference, (iii) second order
structure functions $S_{2,0,0}= \langle (\Delta v_z)^2\rangle$ has
no simple relation with any mixed structure function. This is a
possible indication of $S_{2,0,0}$ having different scaling
properties in the large-$\mathbf\Omega$ limit, a fact we demonstrate
below explicitly in a perturbative calculation by using an {\em
effective model}. Finally, one may eliminate pressure $p^*$ from the
exact hierarchical relations (\ref{voncompact}) and
(\ref{3rdorderII}) by using Eq.~(\ref{press}), although the ensuing
forms of Eqs.~(\ref{3rdorderI}) and (\ref{3rdorderII}) are not
particularly illuminating (we do not report their explicit forms
here). Let us now consider our results as above in the context of
the helical analog of the von Karman-Howarth relation  as studied in
Ref.~\cite{heli}. It may be noted that Ref.~\cite{heli} considers
helical turbulence which is still isotropic, where as rotating
turbulence as considered here not only has finite helicity injected
into the system by the rotation, the system no longer remains
isotropic due to the existence of a preferred direction (the
rotation axis). As a result, Ref.~\cite{heli} has been able to
obtain a closed form explicit analog of the von Karman-Howarth
relation (\ref{vonkar}) for isotropic non-helical $3d$ turbulence.
In contrast, we are able to obtain only a differential form of a von
Karman-Howarth like relation for rotating turbulence. Apart from the
anisotropy effects, yet another notable difference between our
results and those in Ref.~\cite{heli} is the nonlocal nature of the
relations for rotating turbulence. Thus, despite the presence of
helicity in both the cases, our results are not exactly equivalent
of those in Ref.~\cite{heli} in any special limit.

% in the
%large-$\Omega$ limit.
%: We have
%$p^*/\Omega=-\frac{1}{\nabla^2}\nabla\cdot ({\hat z}\times {\bf
%v})\equiv \psi$ in the limit $\Omega\rightarrow\infty$. Substituting
%this in relations (\ref{largeomega}) we find
%\begin{eqnarray}
%\langle \Delta v_\theta\Delta  v_r\rangle&=&-\frac{1}{2} \langle
%(\frac{\partial}{\partial r}\Delta \psi)\Delta v_r\rangle=0,\nonumber\\
%\langle \Delta v_z\Delta v_\theta\rangle &=&\frac{1}{4}[\langle
%\frac{\partial \Delta\psi}{\partial z}\Delta v_r\rangle + \langle
%\frac{\partial \Delta\psi}{\partial r}\Delta v_z\rangle],\nonumber
%\\
%\langle \Delta v_z\Delta v_r\rangle &=&-\frac{1}{4}\langle
%\frac{\partial \Delta \psi}{\partial z}\Delta v_\theta,\nonumber \\
%-\langle (\Delta v_\theta)^2\rangle + \langle (\Delta v_r)^2\rangle
%&=&\frac{1}{4}\langle \frac{\partial \Delta\psi}{\partial r} \Delta
%v_\theta\rangle.\label{largeomegared}
%\end{eqnarray}
%Relations (\ref{largeomegared}) are expressed solely in terms of
%$\bf v$ and hold in the limit of large-$\mathbf\Omega$.

 In the same way as above, analogous {\em exact}
relations between certain third order velocity structure functions
and mixed third order structure functions involving two factors of
velocity differences and one factor of pressure gradient difference
 are obtained in the large
rotation limit. We write some of them below:
\begin{eqnarray}
-2\langle (\Delta v_\theta)^2 \Delta v_r\rangle + \langle (\Delta
v_r)^3\rangle &=& - \frac{1}{2\Omega}\langle (\frac{
\partial}{\partial r_\perp}\Delta p^*)
\Delta v_\theta \Delta v_r\rangle,\nonumber \\
-\langle (\Delta v_\theta)^3\rangle +2\langle (\Delta v_r)^2\Delta
v_\theta\rangle &=&-\frac{1}{4\Omega}\langle
(\frac{\partial}{\partial r_\perp} \Delta p^*) (\Delta
v_\theta)^2\rangle,\nonumber \\
-2\langle (\Delta v_\theta)^2 \Delta v_z\rangle +2 \langle (\Delta
v_r)^2 \Delta v_z\rangle &=&-\frac{1}{4\Omega}[\langle
\frac{\partial\Delta p^*}{\partial z} \Delta v_r \Delta
v_\theta\rangle + \langle \frac{\partial \Delta p^*}{\partial
r_\perp}
\Delta v_\theta \Delta v_z\rangle],\nonumber \\
-2\langle \Delta v_r \Delta v_\theta \Delta v_z\rangle &=&
\frac{1}{4\Omega}[\langle \frac{\partial \Delta p^*}{\partial
r_\perp} (\Delta v_\theta)^2\rangle + 2 \langle \frac{\partial
\Delta p^*}{\partial r_\perp}\Delta v_z \Delta v_r\rangle].
\label{largeomega3}
\end{eqnarray}
 %These as for the second-order structure functions, are {\em exact} relations.
Relations (\ref{largeomega3}) may further be rewritten in terms of
$\bf v$ only by eliminating pressure using Eq.~(\ref{press}). Again,
as for the second order functions, third order function $\langle
(\Delta v_z)^3\rangle$ does not appear in these relations. This
again is an indirect demonstration of anisotropic effect of
rotation. Relations (\ref{2ndorder}) and (\ref{largeomega}) together
constitute a set of exact relations between different second order
structure functions in the limit of $\Omega\rightarrow\infty$.
Similarly, relations (\ref{3rdorderI}), (\ref{3rdorderII}) and
(\ref{largeomega3}) together constitute a complete set
(differential) relations involving different third order structure
functions for rotating turbulence in the large-$\mathbf\Omega$
limit. One may further obtain for any odd positive integer $n$
\begin{equation}
\langle \frac{\partial \Delta p^*}{\partial z} (\Delta
v_z)^{n}\rangle =0 \label{alln}
\end{equation}
for large rotation. Although, these hierarchical relations do not
yield the scaling exponents directly, they serve as benchmarks on
any theoretical (analytical or numerical) and experimental results
on the structure functions. In order to progress further from
Eqs.~(\ref{largeomega}) or (\ref{largeomega3}), one must make
further approximations, see, e.g., Ref.~\cite{sagar1}. Lastly, how
do the relations (\ref{largeomega}) and (\ref{largeomega3}) compare
with those for pure $2d$ turbulence? As shown in
Ref.~\cite{yakhot1}, for pure $2d$ turbulence, pressure does not
contribute to the relations between second order structure functions
or third order structure functions. More generally, the
large-$\Omega$ limit of the steady-state master equation
(\ref{zfinaleqI}) is not the same as that for $d=2$: The master
equation in the large-$\mathbf\Omega$ limit is
\begin{equation}
4 \left(-\eta_1\frac{\partial}{\partial \eta_2} +\eta_2
\frac{\partial}{\partial\eta_1}\right){\mathcal Z}=
I_p/\Omega,\label{eqlargeom}
\end{equation}
as all other terms in Eq.~(\ref{zfinaleqI}) vanish in the
large-$\mathbf\Omega$ limit. The corresponding master equation for
pure $2d$ turbulence in the steady state is given by \cite{yakhot1}
\begin{equation}
\left[\frac{\partial}{\partial r_\perp}\frac{\partial}{\partial
\eta_2}+\frac{2}{r_\perp} \frac{\partial}{\partial\eta_2}
+\frac{\eta_3}{r_\perp} \frac{\partial}{\partial\eta_2}
\frac{\partial}{\partial\eta_3}
-\frac{\eta_2}{r_\perp}\frac{\partial^2}
{\partial\eta_3^2}\right]{\mathcal Z}_2=I_{f2}+I_{p2}+D_2,
\label{2dmaster}
\end{equation}
where ${\mathcal Z}_2$ is the $2d$ analog of Eq.~(\ref{zdefi}),
$I_{f2},\,I_{p2}$ and $D_2$ are the force contributions, pressure
contributions and anomaly terms in $2d$. Clearly,
Eq.~(\ref{zfinaleqI}), the master eqution for $3d$ rotating
turbulence, is different from Eq.~(\ref{2dmaster}), the master
equation for $2d$ non-rotating turbulence. This led us to generally
conclude that the statistical properties of a turbulent rotating
fluid in the large rotation limit is different from pure $2d$
turbulence.

Having discussed anisotropic effects of rotation on the statistical
properties of $3d$ homogeneous turbulence in terms of the
hierarchical relations between structure functions of different
orders, we now set out to calculate the scaling of the equal time
two-point velocity correlation functions (equivalently second order
structure functions) explicitly within a one-loop perturbation
calculation.

\section{Effects of rotation: perturbative analysis}
\label{perturb}

In this Section we use the Navier-Stokes equation (\ref{navier}) in
a rotating frame, subject to stochastic forcings of the type
(\ref{yakhotnoise}). This has a long history in homogeneous and
isotropic fluid turbulence studies which are well-documented in
Refs.~\cite{martin}; see also Ref.~\cite{anisobook} for discussions
on anisotropy and helicity using randomly stirred Navier-Stokes
equation. Despite some well-known limitations and difficulties of
the method it has succeeded in predicting several universal numbers
and scaling exponents which are close to their experimentally
obtained values. Note in the presence of rotation the Coriolis force
introduces anisotropy even at the linear level in the Navier-Stokes
equation [Eq.(\ref{navier})]. Increasing rotation velocity $\bf
\Omega$ (i.e., decreasing Ekman number $M$) is expected to modify
the statistical behavior. In the Fourier space the Navier-Stokes
Eq.~(\ref{navier}) takes the form
\begin{eqnarray}
&&(-i\omega +\nu_0 k^2)v_i + i\frac{\lambda}{2}P_{ijp}({\bf k})
\Sigma_{\bf q} v_j ({\bf q}, \omega_1)v_p ({\bf k-q}, \omega
-\omega_1) -ik_i p^*\nonumber \\ &&= f_i - 2P_{im}({\bf
k})\epsilon_{mjp}\Omega_jv_p({\bf k},\omega), \label{navierF}
\end{eqnarray}
where $\bf k$ is a $3d$ wavevector. Here, density $\rho$ has been
set to unity. The inverse of the bare propagator matrix $G_0^{-1}$
is given by
\begin{eqnarray}
G_0^{-1}=\left[
\begin{array}{ccc}
-i\omega +\nu_0k^2+2P_{xy}\epsilon_{yzx}\Omega &
2P_{xx}\epsilon_{xzy}\Omega & 0\\
2P_{yy}\epsilon_{yzx}\Omega &
-i\omega+\nu_0k^2+2P_{yx}\epsilon_{xzy}\Omega & 0\\
2P_{zy}\epsilon_{yzx}\Omega & 2P_{zx}\epsilon_{xzy}\Omega & -
i\omega +\nu_0 k^2
\end{array}\right].
\label{bareprop}
\end{eqnarray}
Here $\epsilon_{ijk},\,i,j,k=x,y$ or $z$, is the totally
antisymmetric tensor in $3d$: $\epsilon_{xyz}=1$ etc. Different
elements of $G_0$ are given in Sec.~\ref{appenII}. The renormalized
correlation matrix $C_{ij}({\bf k},\omega)$ is formally given by
\begin{equation}
C_{ij}({\bf k},\omega)\equiv \langle v_i ({\bf k},\omega) v_j({\bf
-k},-\omega\rangle = G_{im}({\bf k},\omega) \langle f_m({\bf
k},\omega )f_p({\bf -k},-\omega)\rangle G_{pj}({\bf -k},-\omega),
\label{corrdef}
\end{equation}
where $G_{ij}({\bf k},\omega)$ is the fully renormalized version of
the bare propagator $G_{ij0}({\bf k},\omega)$ given by the matrix
(\ref{bareprop}).

To find out non-linear effects of rotation and the renormalized
correlation function, we now need to find out the fluctuation
corrections to the different elements of $G_0$ in a systematic
perturbation theory. It is evident that any straight forward
perturbation theory will be much more algebraically complicated due
to the non-isotropic nature of $G_0({\bf k},\omega)$, reflected by
the last term in the right hand side of Eq.~(\ref{navierF}), than
its isotropic counter part~\cite{yakhot}. In order to circumvent
this algebraic difficulty we use a {\em modified} equation as an
{\em effective} equation to calculate the renormalization of the
elements of $G_0$. We replace, in the last term of
Eq.~\ref{navierF}, $v_p ({\bf k},\omega)$ by $\hat G_0({\bf
k},\omega)f_p({\bf k},\omega)$: Thus we obtain
\begin{eqnarray}
&&\hat G_0^{-1}v_i + i\frac{\lambda}{2}P_{ijp}({\bf k}) \Sigma_{\bf
q} v_j ({\bf
q}, \omega_1)v_p ({\bf k-q}, \omega -\omega_1) -ik_i p^*\nonumber \\
&&= f_i - 2\epsilon_{ijp}\Omega_j\hat G_0({\bf k},\omega)f_p({\bf
k},\omega). \label{model}
\end{eqnarray}
Here, $\hat G_0$ is the bare propagator of Eq.~(\ref{model}):
\begin{equation}
\hat G_0({\bf k},\omega)=\frac{1}{-i\omega + \nu_0k^2}.
\end{equation}
Further the effective pressure may be eliminated by using the
incompressibility constraint $\nabla\cdot {\bf v}=0$. We obtain
\begin{eqnarray}
&&(-i\omega +\nu_0 k^2)v_i + i\frac{\lambda}{2}P_{ijp}({\bf k})
\Sigma_{\bf q} v_j ({\bf q}, \omega_1)v_p ({\bf k-q}, \omega
-\omega_1)\nonumber \\ &&= f_i - 2P_{im}({\bf k})\epsilon_{mjp}\Omega_j\hat G_0({\bf k},\omega)f_p({\bf k},\omega)\nonumber \\
\equiv \phi_i ({\bf k},\omega).
\label{basic}
\end{eqnarray}
The correlation of the effective noise $\phi_i$ is zero-mean, Gaussian distributed with a variance
\begin{eqnarray}
&&\langle \phi_i ({\bf k},\omega) \phi_j ({\bf -k},-\omega)\rangle
\nonumber \\ &&= 2D_0 |k^{-y}|P_{ij} - 4P_{jm}\epsilon_{mnp}\hat
G_0(-{\bf k},-\omega) P_{ip} ({\bf k})|k^{-y}|\nonumber \\
&&-4P_{im}({\bf k}\epsilon_{mnp}\Omega_n\hat G_0({\bf
k},\omega)P_{jp}({\bf k})D_0|k|^{-y} \nonumber \\ &&+ 8D_0
P_{im}P_{js}\epsilon_{mnp}\epsilon_{srq}\Omega_n\Omega_r |\hat
G_0({\bf k},\omega)|^2 P_{pq}({\bf k})|k|^{-y}. \label{noisecorr}
\end{eqnarray}
This {\em effective model}, although not exact, retains two basic
effects of rotation, namely, anisotropicity and breakdown of parity
invariance. We now use this effective noise with a variance
(\ref{noisecorr}) together with the {\em effective model} in the
incompressible limit (\ref{basic}) to calculate the inertial range
of the two-point equal time correlations of different velocity
components. The calculational advantage of the effective model
Eq.~(\ref{model}) stems from the fact that its bare propagator is
identical to $\hat G_0$, i.e., isotropic. Below we use
Eq.~(\ref{basic}) as a starting point for our analyses below.

Formally, fluctuation corrections to the bare propagator is given by
the Dyson equation
\begin{equation}
\hat G^{-1}_{\alpha\beta}({\bf k},\omega) = \hat
G_0^{-1}\delta_{\alpha\beta} ({\bf k},\omega) -\Sigma_{\alpha\beta}
({\bf k}, \omega), \label{dyson}
\end{equation}
where $\hat G_{\alpha\beta}({\bf k},\omega)$ is the {\em
renormalized} (fluctuation corrected) propagator, $\hat G_0({\bf
k},\omega)$ is the bare propagator and $\Sigma_{\alpha\beta} ({\bf
k},\omega)$ is self-energy which arises due to the non-linear terms
and contain fluctuation corrections.
%\begin{figure}[htb]
%\hskip-1cm
%\includegraphics[height=4.5cm]{prop.eps}\hspace{2cm}
%\caption{The one-loop diagram which renormalizes the bare
%propagator. A line with a filled circle indicates a correlator and a
%line indicates a propagator.} \label{fig1}
%\end{figure}
The self-energy at the one-loop level is given by (in terms of the
bare propagator $\hat G_0 ({\bf k},\omega)$)
\begin{eqnarray}
\Sigma_{ls}&=&-\lambda^2P_{lmn}({\bf
k})\int\frac{d\omega_1}{2\pi}\frac{d^dq}{(2\pi)^d}\frac{1}{\omega_1^2
+\nu_0 q^4}\frac{1}{-i\omega_1 + \nu_0 ({\bf k-q})^2}\nonumber \\&&
\times P_{nps}({\bf k-q}) \langle \phi_m ({\bf q},\omega_1) \phi_p
({\bf -q},-\omega_1)\rangle \nonumber \\
&&=-\frac{\lambda^2}{2}P_{lmn}({\bf k})\int\frac{d\omega_1}{2\pi}
\frac{d^dq}{(2\pi)^d}\frac{P_{nps}({\bf k-q})}{\omega_1^2 +\nu_0^2
k^4}\nonumber \\&& \frac{2D_0 |q|^{-y}}{-i\omega_1 + \nu_0 ({\bf
k-q})^2}[ P_{mp}({\bf q}) + 2 P_{p\alpha} ({\bf q}) P_{m\beta}\Omega
\epsilon_{\alpha z\beta}\nonumber \\&&\frac{2i\omega_1}{\omega_1^2
+\nu_0 q^4}+ 4P_{m\alpha}({\bf q}) P_{p\beta}\epsilon_{\alpha z
\nu}\epsilon_{\beta z \delta}\Omega^2
\frac{P_{\nu\delta}}{\omega_1^2 +\nu_0q^4}]. \label{sigma}
\end{eqnarray}

 Expression
(\ref{sigma}) shows that the self-energy $\Sigma_{ls}$ has an
$\Omega$-independent part which is identical to the one-loop
self-energy expression for the non-rotating case, an $O(\Omega)$
part and an $O(\Omega^2)$ part: The $O(\Omega)$ contributions to
different elements of $\Sigma_{ls}$ are (for details see
Sec.~\ref{appenIII})
\begin{eqnarray}
%\Sigma_{zz}&=&0,\nonumber \\
\Sigma{ij}({\bf k},\omega=0)&\sim&
P_{il}\epsilon_{jzl}\frac{Dk^2\Omega}{\nu_0^3}\int dq q^{d-1}
q^{-y-6}, \label{oneloopsigma}
\end{eqnarray}
where Cartesian indices $i,\,j$ refer to $x,\,y$ or $z$, such that
the first order solution to the velocity field is given by
$\Sigma_{il} \epsilon_{jzl}\phi_j$. It is clear that there are no
$O(\Omega)$ corrections to $\Sigma_{zz}$. Furthermore, the
zero-frequency one-loop contribution to $\Sigma_{ij}$ is real. The
$O(\Omega^0)$ contribution to $\Sigma_{zz}$ has infra-red
divergence. Demanding self-consistency and noting that there are no
fluctuation corrections to the noise strength $D_0$ \cite{yakhot},
we find that the scale-dependent renormalized viscosity $\nu (k)\sim
\nu k^{-4/3}$, where we have ignored effects of anisotropy. Noting
that in the long wavelength limit, the {\em effective} or {\em
renormalized} viscosity diverges as $\nu k^{-4/3}$ \cite{yakhot},
where $\nu$ is a (dimensional) numerical coefficient, and using it
in place of bare viscosity $\nu_0$ in (\ref{oneloopsigma}) we obtain
\begin{equation}
\Sigma_{ij}(k)\sim P_{il}\epsilon_{jzl}k^{2/3} M^{-1},
\end{equation}
as the $O(\Omega)$ correction to the self-energy. Here,
$M^{-1}=2\Omega/(\nu k^{2/3})$  is the {\em renormalized} or {\em
scale-dependent} inverse Ekman number.  Thus the $\Omega$-dependent
corrections will appear as a series in $M^{-1}(k)$. With this
scale-dependent self-energy $\Sigma_{ij}(k)$ we now calculate the
different elements of the two-point velocity correlation function
matrix. While doing this we use the effective equation (\ref{basic})
together with the noise variance (\ref{noisecorr}) which now depends
upon the scale-dependent propagator $G_{ij}$. As for the effective
noise, since we have
\begin{eqnarray}
%\phi_z({\bf k},\omega)&=& f_z({\bf k},\omega),\nonumber \\
\phi_i({\bf k},\omega)&=& f_i({\bf k},\omega) -
2P_{im}\epsilon_{mzp}G_{ps}f_s
\end{eqnarray}
with $i=x,y,z$, the $\Omega$-dependent part of the effective noise
$\phi_j$ has no contribution from $G_{zz}({\bf k},\omega)$.
Explicitly calculating we find
\begin{eqnarray}
G_{11}&=&\frac{-i\omega + \nu k^{2/3} - 2P_{xy}\Omega}{(-i\omega +
\nu k^{2/3})^2 + 4\frac{k_z^2}{k^2}\Omega^2},\nonumber \\
G_{12}&=& \frac{2P_{yy}\Omega}{(-i\omega + \nu k^{2/3})^2
+4\frac{k_z^2}{k^2}\Omega^2},\nonumber \\
G_{13}&=&\frac{4P_{yy}P_{zx}\Omega^2 + 2P_{zy}(-i\omega + \nu
k^{2/3} - 2P_{xy}\Omega)\Omega}{[(-i\omega + \nu k^{2/3})^2
+4\frac{k_z^2}{k^2}\Omega^2](-i\omega + \nu k^{2/3})},\nonumber \\
G_{21}&=&\frac{2P_{xx}\Omega}{(-i\omega + \nu k^{2/3})^2
+4\frac{k_z^2}{k^2}\Omega^2},\nonumber \\
G_{22}&=&\frac{-i\omega + \nu k^{2/3} + 2P_{xy}\Omega}{(-i\omega +
\nu k^{2/3})^2 + 4\frac{k_z^2}{k^2}\Omega^2},\nonumber \\
G_{23}&=&\frac{4P_{xx}P_{zy}\Omega^2 - 2P_{zx}\Omega (-i\omega + \nu
k^{2/3} - 2P_{xy}\Omega)}{[(-i\omega + \nu k^{2/3})^2 +
4\frac{k_z^2}{k^2}\Omega^2]},\nonumber \\
G_{33}&=&\frac{1}{-i\omega + \nu k^{2/3}},\nonumber \\
G_{31}&=&0=G_{32}.
\end{eqnarray}
Let us now consider various correlation functions. We find
\begin{eqnarray}
C_{zz}({\bf k},\omega) &=& \langle |v_z({\bf
k},\omega)|^2\rangle=\frac{\langle \phi_z ({\bf
k},\omega)\phi_z({\bf -k},-\omega)\rangle}{\omega^2 +\nu^2
k^{4/3}}.%\nonumber \\ &=& \frac {2D_0 k^{-3}}{\omega^2 + \nu^2
%k^{4/3}}[P_{zz} + ...].
\end{eqnarray}
Thus, the equal time correlator $C_{zz}(k,t=0)\sim k^{-11/3}$ in the
long wavelength limit. In contrast, all other correlators scale
differently in the inertial range. Let us consider one of them,
namely, $C_{xx}=\langle |v_x({\bf k},\omega)|^2\rangle$. We begin by
noting that
\begin{equation}
v_x=G_{11}\phi_x + G_{12}\phi_y + G_{13}\phi_z.\label{vx}
\end{equation}
Correlation $C_{xx}$ may now be calculated from (\ref{vx}) by using
the effective noise variance. A full calculation keeping the proper
anisotropicity is algebraically difficult; hence we perform a
scaling level calculation ignoring anisotropic amplitudes. Keeping
in mind that the propagators that appear in the noise variance
should be renormalized propagators, we find that, since $\phi_i$
does not receive any contribution from $G_{zz}$, we find
\begin{eqnarray}
C_{xx}({\bf k},\omega)=\langle | G_{11}\phi_x + G_{12}\phi_y +
G_{13}\phi_z|^2\rangle.
\end{eqnarray}
Noting that the effective noise $\phi_i$ involves all elements of
the propagator matrix except for $G_{zz}$, the leading scale
dependence of $C_{xx}$ is given by $C_{xx}\sim k^{-3}$ to the
leading order in $\Omega^2$ (again we have ignored the anisotropy).
In a similar way, $C_{yy}$ also scales as $k^{-3}$. We have already
shown above that $C_{zz}$ scales as $k^{-11/3}$. Thus $C_{xx}$ and
$C_{yy}$ scales very differently from $C_{zz}$. The latter scales
same as in $3d$, where as $C_{xx}$ and $C_{yy}\sim k^{-3}$ is less
steep than $C_{zz}$. For each of these we however only extract the
scale-dependence and no result on the dependences on $k_z$ and $k$
separately. Let us see what our results above mean for
one-dimensional energy spectra. Since $\langle |v_z({\bf
k},t)^2\rangle$ scales differently from $\langle |v_j({\bf
k},t)|^2\rangle,\,j=x,y$, it is natural to define two
one-dimensional energy spectra: $E_\parallel (k)\sim k^2 \langle
|v_z({\bf k},t)^2\rangle \sim k^{-5/3}$ and $E_\perp \sim k^2
\langle |v_j({\bf k},t)|^2\rangle\sim k^{-1},\,j=x,y$, where the
factor $k^2$ appears due to the isotropic phase factor in $3d$
(since we ignored the anisotropic coefficient we have used the $3d$
isotropic phase factor). As our scaling level results do not
distinguish between different directions in the Fourier space, the
energy spectra obtained above also do not have any anisotropy.
Finally, in a related issue, our results here provide for a
naturally occurring example of a situation where scaling are
different in different directions, see, e.g., Ref.~\cite{anisocrit}
and references therein for examples of models exhibiting such
behavior. Let us now compare with the existing results: Recent
direct numerical studies \cite{mueller} of the forced Navier-Stokes
equation (\ref{navier}) for a rotating fluid for large rotation
yield for one-dimensional spectrum $E({\bf k_\perp})\sim
k_\perp^{-2}$, where $\bf k_\perp$ is perpendicular to
$\mathbf\Omega$. Similarly numerical solutions of shell model
equations \cite{sagar} for rotating turbulence yield for energy
spectrum $E(k)\sim k^{-2}$ for large rotation (shell models, being
one-dimensional models, do not distinguish between different
directions). Weak turbulence theory \cite{galtier} approach as well
suggests a spectrum $\sim k^{-2}_\perp$. In contrast, our one-loop
mode-coupling calculations for {\em small}-$\Omega$ yields spectra
$k^{-1}$, less steep than those obtained in earlier numerical
approaches. However, two things should be kept in mind while
comparing our results with results obtained from other approaches:
(i) our perturbative calculation and the effective equation are
valid only for small-$\Omega$, (ii) we did not distinguish between
different directions (due to intractable and complicated nature of
the underlying one-loop integrals), and our consequent usage of the
$3d$ phase factor is actually not appropriate. An anisotropic phase
factor is likely to change the scaling of energy spectra here. These
make direct comparisons with other existing results difficult.

\section{Conclusion}
\label{conclu}

In this article we have analyzed the effects of rotation on the
scaling properties of $3d$ homogeneous incompressible turbulence. We
have used a two-pronged strategy: First of all, we set up
hierarchical relations between structure functions of different
orders by using an approach used in Refs.~\cite{yakhot1,sreeni}.
Unlike the isotropic case, there are no closed relations involving
the third-order structure functions. Moreover, second-order
structure functions appear in those relations. Thus simple analogs
of the well-known von Karman-Howarth relation for isotropic fluid
turbulence do not exist here. Furthermore, mixed second-order
structure functions made of pressure gradients and differences of
velocity components appear in these relations. We are able obtain a
differential form analog of the von Karman-Howarth relation of
non-rotating fluid turbulence. However, this is non-integrable. All
these features are in contrast to the results for isotropic
turbulence (${\mathbf\Omega}=0$). In the limit of
large-$\mathbf\Omega$ these relations yield exact relations between
certain second-order velocity structure functions and mixed second
order structure functions of pressure gradient and differences of
velocity components. The overall structures of the hierarchical
relations suggest that the scaling properties of velocity structure
functions involving one or more in-plane (plane perpendicular to the
axis of rotation) is expected to be very different from
$S_{2,0,0}=\langle (\Delta v_z)^2\rangle$. In a similar way, we are
able to derive exact relations between third-order velocity
structure functions and mixed third-order structure functions
involving two factors of velocity differences and one factor of
pressure gradient difference in the limit of large rotation. Again,
similar to the case of the second-order structure functions
third-order structure function $S_{3,0,0}\equiv \langle (\Delta
v_z)^3\rangle$ does not appear in any of these relations,
reinforcing further the effects of anisotropy due to rotation. We
then ask: Are the turbulent flows at $\Omega\rightarrow\infty$
statistically same as pure $2d$ turbulence? As we have discussed
above, the two are {\em not} identical, a fact which is clearly
brought out by the respective structure function hierarchies in the
two cases. Thus our conclusions are in agreement with that of
Ref.~\cite{galtier}. With these exact relations at hand, we then
embark upon an explicit calculation of the scaling of different
two-point equal time velocity correlation functions. In order to
simplify the ensuing calculations we use an effective model and use
a one-loop approximation for our purposes. The resulting one-loop
integrals are complicated due to the anisotropic nature of the
system. Treating them at the scaling level, i.e., ignoring
anisotropy, we obtain $\langle |v_z ({\bf k},t)|^2\rangle \sim
k^{-11/3}$, identical to the result in three dimensions. In contrast
all other two-point equal time velocity correlators display a
scaling $\sim k^{-3}$. The latter result imply a three-dimensional
spectra (again at the scaling level) $\sim k^{-1}$, different from
the conclusions arrived at by using other methods for large
rotation. We would like to emphasize that our mode-coupling results
are only at the scaling level and hold for small rotation; hence
these are only illustrative of the anisotropic scaling due to
rotation. In order to compare with numerical results or results
obtained by other analytical means, more elaborate calculations,
keeping the anisotropic coefficients of the one-loop integrals,
should be performed. Our results, especially on the relations
between different structure functions, may be tested by direct
numerical simulations or experiments: One needs to
calculate/measure, e.g., the two sides of the relations
(\ref{3rdorderI}) or (\ref{3rdorderII}, and determine their
validity.

Our hierarchical relations are {\em exact}; however they are not
closed and cannot be solved. Despite that they bring out two
important results (albeit indirectly): (i) scaling of the two-point
structure function $S_{2,0,0}$ is likely to be different from the
scaling of all other two-point structure function,  (ii) statistical
properties of rotating turbulent flows are not the same as those of
pure $2d$ flows. In contrast, our perturbative calculations are
approximate, but they still provide explicit results on the scaling
properties of the various two-point velocity correlation functions
which are not in contradiction with the exact hierarchical
relations. One-loop diagrammatic calculations presented here suffer
from several limitations which are well-documented in standard
literature. Despite these difficulties we are able to obtain useful
results from them and open up many new relevant and interesting
questions for future studies. As a next step, it would be useful to
account for the anisotropy in the one-loop integrals and find out
the anisotropic scaling functions in the expressions of various
correlators. Further, it would be interesting to find out
perturbatively or numerically whether large rotation may lead to
unequal dynamic exponents for different correlation functions. If
that happens then it would be a natural example of what is known as
{\em weak dynamical scaling} in the literature. Until now the latter
has been observed only in simple model studies
\cite{abhik-mustansir}. Finally some shell model studies
\cite{sagar} indicated that as $\Omega$ increases intermittency
corrections to the Kolmogorov's simple scaling exponents decrease
and finally disappear for high $\Omega$. Given the fact that shell
models do not distinguish between various directions, it would be
useful to address this question by using Direct Numerical
Simulations (DNS) of the $3d$ Navier Stokes equation in a rotating
frame and see whether multiscaling disappears only for the structure
functions made of in-plane velocity components and survives for the
direction parallel to $\mathbf \Omega$. In general our results
suggest that experiments or DNS studies should measure the scaling
of structure functions $\langle (\Delta v_j)^n\rangle$ for
$j=z,r,\theta$ with positive $n$. It should be tested whether
$\langle (\Delta v_z)^n\rangle$ shows simple scaling for
large-$\mathbf \Omega$ for all $n>0$. We hope that our results will
motivate more detailed experimental work as well in the directions
as discussed above.

\section{acknowledgement}
One of us (AB) wishes to acknowledge the Max-Planck-Society
(Germany) and the Department of Science and Technology (India) for
partial financial support for research through the Partner Group
program (2009).
\section{Appendix I: Calculation of the dissipative anomaly}
\label{appenI}

We present explicit evaluation of dissipative anomaly $D$: We begin
with
\begin{eqnarray}
&&\langle \nu_0 \Sigma_i \partial_i^2 \exp [\lambda_z \Delta v_z +
\eta_1 \Delta v_r +\eta_2 \Delta v_\theta]\rangle\nonumber
\\&=& \nu_0 \langle \lambda_z [\nabla_2^2 v_z ({\bf x}_2) -\nabla_1^2
v_z({\bf x}_1)]{\mathcal z}_0\rangle + \nu_0 \langle \eta_1
[\nabla_2^2 v_r ({\bf x}_2)- \nabla_1^2 v_r] {\mathcal Z}_0\rangle +
\nu_0 \langle \eta_2[\nabla_2^2 v_\theta ({\bf x}_2) - \nabla_1^2
v_\theta ({\bf x}_1)]{\mathcal Z}_0\rangle \nonumber \\ &+& 2\nu_0[
\langle \lambda_z^2 (\nabla_i v_z)^2 \rangle + \langle \eta_1
(\nabla_i v_r)^2\rangle + \langle \eta_2 (\nabla_i
v\theta)^2\rangle\nonumber \\ &+& 2\lambda_z\eta_1 \langle(\nabla_i
v_z)(\nabla_i v_r)\rangle +2\eta_1\eta_2\langle (\nabla_i
v_r)(\nabla_i v_\theta) +2\eta_2\lambda_z (\nabla_i
v_\theta)(\nabla_i v_z)\rangle]{\mathcal Z}_0\rangle.\label{dissi1}
\end{eqnarray}
The left hand side of Eq.~(\ref{dissi1}), upon expansion yields
various structure functions of different order multiplied by a
factor of $\nu_0$. Since structure functions themselves are finite
in the limit $\nu_0\rightarrow 0$, the left hand side of
(\ref{dissi1}) vanishes in the limit $\nu_0\rightarrow 0$. Thus we
obtain
\begin{eqnarray}
D&\equiv& \nu_0 \langle \lambda_z [\nabla_2^2 v_z ({\bf x}_2)
-\nabla_1^2 v_z({\bf x}_1)]{\mathcal z}_0\rangle + \nu \langle
\eta_1 [\nabla_2^2 v_r ({\bf x}_2)- \nabla_1^2 v_r] {\mathcal
Z}_0\rangle + \nu \langle \eta_2[\nabla_2^2 v_\theta ({\bf x}_2) -
\nabla_1^2 v_\theta ({\bf x}_1)]{\mathcal Z}_0\rangle \nonumber \\
&-&2 [\langle \lambda_z^2 \epsilon_z \rangle + \langle \eta_1
\epsilon_r \rangle +  \langle \eta_2 \epsilon_\theta \rangle +
2\lambda_z\eta_1\nu(\nabla_i v_z)(\nabla_i v_r) +2\eta_1\eta_2\nu
(\nabla_i v_r)(\nabla_i v_\theta) \nonumber \\&+&2\eta_2\lambda_z\nu
(\nabla_i v_\theta)(\nabla_i v_z)]{\mathcal Z}_0\rangle,
\end{eqnarray}
with $\epsilon_z=(\nabla_i v_z)^2,\,\epsilon_r = (\nabla_i
v_r)^2,\,\epsilon_\theta = (\nabla_i v_\theta)^2$.

%\section{Hierarchical relations between structure functions}
%abel{hierar}
%In this Section we set up the exact hierarchical equations between
%e two-point structure functions of various orders. These equations are however
%t closed. These exact relations, nevertheless, show the differences between
%e rotating and the non-rotating cases.
%In order to find the hierarchy of equations for the equal-time structure functions in the statistical stationary state of the
%ochastically forced Navier-Stokes equation in  a rotating frame for the velocity field $v_i$:
%egin{eqnarray}
%mathcal Z}({\mathbf \lambda}_1, {\mathbf \lambda}_2, {\bf r}_1, {\bf r}_2, t)\equiv
%angle \exp ({\mathbf \lambda}_1\cdot {\bf u}({\bf x}_1) + {\mathbf \lambda}_1\cdot {\bf b}({\bf x}_2)\rangle\nonumber \\
%=\int du_1 du_2{\mathcal P} ({\bf u}_1, {\bf u}_2, {\bf r}_1, {\bf r}_2,t) \exp ({\mathbf \lambda}_1\cdot {\bf u}({\bf x}_1) +
%mathbf \lambda}_1\cdot {\bf b}({\bf x}_2)
%nd{eqnarray}
%ere
\section{Appendix II: Effective noise variance}
\label{appenII} By definition
\begin{eqnarray}
&&\langle \phi_i ({\bf k},\omega) \phi_j ({\bf -k},-\omega)\rangle =
\langle f_i ({\bf k},\omega) f_j({\bf -k},-\omega)\rangle\nonumber
\\ &&
- \langle f_i ({\bf k},\omega)2P_{im} ({\bf
k})\epsilon_{mnp}\Omega_nG_0({\bf -k},-\omega) f_p ({\bf
-k},-\omega)\rangle \nonumber \\ && -\langle 2P_{im}({\bf
k})\epsilon_{mnp}\Omega_n G_0 ({\bf k},\omega)f_p({\bf
k},\omega)f_j({\bf -k},-\omega)\rangle\nonumber \\ && + \langle
2P_{im}\epsilon_{mnp}\Omega_n G_0({\bf k},\omega)f_p ({\bf
k},\omega)\nonumber \\ &&\times 2P_{js}({\bf k})
\epsilon_{srq}\Omega_r G_0({\bf -k},\omega)f_q({\bf
-k},-\omega)\rangle.
\end{eqnarray}
Substituting from Eq.~(\ref{yakhotnoise}) we obtain
Eq.~(\ref{noisecorr}).

\section{Appendix III} \label{appenIII}

Here we write the full expressions of the $\Omega$-independent $I_0$
and $0(\Omega)$ part $I_\Omega$:
\begin{eqnarray}
I_0&=& P_{lmn}({\bf k})\int \frac{d^3q}{(2\pi)^3}\frac{2D_0
P_{mp}({\bf q}) O_{nps} ({\bf q}) |q|^{-y}}{2\nu_0^2 q^2[q^2 + ({\bf
k-q})^2]},\label{I0}\\
I_\Omega &=& \frac{\lambda^2}{2} P_{lmn}({\bf k}) \int\frac{d^3
q}{(2\pi)^3} 4D_0 \Omega|q|^{-y}P_{nps}({\bf k-q})P_{p\alpha}({\bf
q})P_{m\beta} ({\bf q})\epsilon_{\alpha z\beta}\nonumber \\ &\times&
\left[ -\frac{3}{2\nu_0^2 q^2[q^2 + ({\bf k-q})^2]}\right].
\label{I1}
\end{eqnarray}
The above integrals are formally divergent as the external
wavevector ${\bf k}\rightarrow 0$. Self-consistency is achieved for
a scale-dependent viscosity $\nu(k)\nu k^{-4/3}$ and scale-dependent
inverse Ekman number $M(k)\sim \nu k^{2/3}/2\Omega$.

\end{document}